\documentclass[12pt,twoside]{article}

\usepackage{arxiv}
\usepackage{amsmath}
\usepackage{amssymb}
\usepackage{amsfonts}
\usepackage{natbib}
\usepackage{graphicx}
\usepackage{placeins}
\usepackage{array}
\usepackage{multirow}
\usepackage{multicol}
\usepackage{dcolumn}
\usepackage{mathtools}
\usepackage{enumitem}

\usepackage[american]{babel}
\usepackage{hyphenat}

\newtheorem{example}{Example}[section]
\newtheorem*{notation}{Notation}

\theoremstyle{plain}
\newtheorem{algorithm}{Algorithm}

\newtheorem{condition}{Condition}

\numberwithin{equation}{section}

\let\oldemph\emph
\renewcommand{\emph}[1]{\textcolor{blue!60!black}{\oldemph{#1}}}

\usepackage{hyperref}
\usepackage{cleveref}
\crefname{figure}{figure}{figures}
\hypersetup{
		pdfencoding=auto, 
		psdextra,
		colorlinks=true,
		citecolor=green!40!black,
		linkcolor=red!50!black,
		urlcolor=blue!80!black
	}

\DeclareMathOperator{\Unif}{Unif}

\usepackage{todonotes}

\input{tcilatex}
\newcolumntype{R}{>{\flushright\arraybackslash}b{.25in}}

\newcommand{\rel}[1]{\hat{#1}}
\newcommand{\all}[1]{#1}
\newcommand{\floor}[1]{\left\lfloor{#1}\right\rfloor}
\newcommand{\lfrac}[2]{{#1}/{#2}}

\ORDtitle{Seat Allocation and Seat Bias under the Jefferson--D'Hondt Method}
\ORDshorttitle{Seat Allocation and Seat Bias...}
\ORDshortafill{D. Boratyn et al.}
\ORDaffilone{Center for Quantitative Political Science, Jagiellonian University, Krakow, Poland}
\ORDaffiltwo{Faculty of Mathematics and Computer Science, Jagiellonian University, Krakow, Poland}
\ORDaffilthree{Faculty of International and Political Studies, Jagiellonian University, Krakow, Poland}
\ORDaffilfour{}
\ORDauthorone{Daria~Boratyn}{1}{0000-0003-3299-7071}{daria.boratyn@uj.edu.pl}
\ORDauthortwo{Wojciech~Słomczyński}{1,2}{0000-0003-2388-8930}{wojciech.slomczynski@uj.edu.pl}
\ORDauthorthree{Dariusz~Stolicki}{1,3}{0000-0002-8295-0848}{dariusz.stolicki@uj.edu.pl}
\ORDauthorfour{}{}{}{}
\ORDauthorfive{}{}{}{}
\ORDcorrespoding{dariusz.stolicki@uj.edu.pl}

\ORDnumber{0}
\ORDvolume{0}
\ORDdoi{DOI: draft}
\ORDyear{2024}
\ORDReceived{xxxxx~xxxxx}
\ORDRevised{xxxxx~xxxxx}
\ORDAccepted{xxxxx~xxxxx}
\ORDPublished{xxxxx~xxxxx}
\ORDFirstpagenum{1} 

\renewcommand{\citeyear}[1][]{\cite{#1}}
\renewcommand{\citeyear}[2][]{\cite{#1,#2}}

\begin{document}

\maketitle

\begin{abstract}
We prove that under the Jefferson--D'Hondt method of apportionment, given certain distributional assumptions regarding mean rounding residuals, as well as absence of correlations between party vote shares, district sizes (in votes), and multipliers, the seat share of each relevant party is an affine function of the aggregate vote share, the number of relevant parties, and the mean district magnitude. We further show that the first of those assumptions follows approximately from more general ones regarding smoothness, vanishing at the extremes, and total variation of the density of the distribution of vote shares. 
We also discuss how our main result differs from the simple generalization of the single-district asymptotic seat bias formulae, and how it can be used to derive an estimate of the natural threshold and certain properties thereof.
\end{abstract}

\ORDKeywords{apportionment, Jefferson--D'Hondt method, seat bias, rounding}


\section{\label{secIntro}Introduction}

The \textbf{Jefferson--D'Hondt method of apportionment}\footnote{The Jefferson--D'Hondt method is also known as the Hagenbach-Bischoff method \citep[p.~204]{Szpiro10}, the method of greatest divisors \citep{Huntington21, Huntington28, Huntington31}, the method of highest averages \cite[pp.~17-19]{Carstairs80}, and the method of rejected fractions \citep{Chafee29}. In Israel the method is called the Bader-Ofer method after two members of the Knesset who proposed it in 1975: Yohanan Bader and Avraham Ofer.} was originally devised in 1792 by Thomas Jefferson to apportion seats in the U.S. House of Representatives among the states \citep{Jefferson92}, and later it was proposed by a Belgian mathematician and lawyer Victor D'Hondt \citeyear{DHondt82, DHondt85}\footnote{\citet{Dancisin13a} discusses the evolution of D'Hondt's ideas on the subject of proportional representation and the origins of his method.
The method was also rediscovered by several authors in various contexts between 1860 and 1874, see \citet[p.~6]{MoraGine13} for details.} for use in parliamentary elections\footnote{It is unclear whether D'Hondt knew of Jefferson's work on the subject. \citet[p.~36]{James97} has probably been the first to notice that the Jefferson method is equivalent to the D'Hondt method, but it appears that this finding has escaped the attention of the subsequent generations of scholars. As far as we are aware, \citet[p.~703]{BalinskiYoung75} have been the first modern authors to credit Jefferson with the original authorship of the method. It should be noted that the Jefferson method was enacted into law (Act of Apr. 14, 1792, c. 23, 1 Stat. 253) and had remained in use for apportioning representatives among the states until 1842. For an extensive discussion of the history of congressional apportionment methods, see \citet{Biles17}.}.
It calls for finding such a divisor $\all{\delta}$ that if each party $i=1,\dots ,\all{n}$, where $\all{n} \in \mathbb{N}$ is the number of parties, were to be allocated as many seats $s_{i}$ as its number of votes $v_{i}$ divided by $\all{\delta}$, rounding down to the nearest integer, i.e., if $s_{i}=\floor{ v_{i}/\all{\delta}}$, no seats would remain unallocated, i.e., $\sum_{i=1}^{\all{n}} s_{i}=s$, where $s$ is the overall number of seats. It is easy to demonstrate that -- unless an electoral tie occurs -- there are always uncountably many such divisors and they always yield the same apportionment of seats.

Whether because of its relative simplicity or a bias in favor of the largest parties (who are usually setting the rules), the Jefferson--D'Hondt method has attained wide popularity. It is currently employed to allocate all or some parliamentary seats in Albania, Argentina, Aruba, Austria, Belgium, Cape Verde, Chile, Croatia, the Czech Republic, Denmark, the Dominican Republic, East Timor, Faroe Islands, Fiji, Finland, Greenland, Iceland, Israel, Japan, Luxembourg, Macedonia, Montenegro, the Netherlands, Paraguay, Peru, Poland, Portugal, S\u{a}o Tome and Pr\'{\i}ncipe, Serbia, Spain, Suriname, Switzerland, and Turkey \citep{FlisEtAl20}, and to allocate European Parliamentary seats in a majority of the EU member states, making it one of the most popular proportional representation formulae \citep[see, e.g.,][]{BormannGolder13, Carey17, Colomer04}.

Drawing on earlier works by \citet{Janson14} and \citet{Pukelsheim14} (see Sec.~\ref{secLit}), 
\citet{FlisEtAl20} have proposed a \textbf{seat allocation formula} describing, under assumptions discussed in Sec.~\ref{secProof}, the relationship between the seat share of the $i$-th party $q_{i}$ and the vector of aggregate electoral results $\left( v_{1},\dots ,v_{\all{n}} \right) $, given mean district magnitude $m$ and assuming that parties are sorted degressively by the number of votes
\begin{equation}
q_{i}=\begin{dcases} \rel{p}_{i}+\rel{p}_{i}\frac{\rel{n}}{2m}-\frac{1}{2m} & \textrm{for } i \le \rel{n},\\
0 & \textrm{for } i > \rel{n}, \end{dcases} \label{eq:seatAlloc}
\end{equation}
where $\rel{n}$ is the number of ``relevant'' parties
\begin{equation}
\rel{n}:=\max \left\{ l=1,\dots,\all{n}:\frac{v_{l}}{\sum_{j=1}^{l}v_{j}}>\frac{1}{2m+l}\right\} ,  \label{eq:relevParties}
\end{equation}
and $\rel{p}_{i}$ is the renormalized vote share of the $i$-th party
\begin{equation}
\rel{p}_{i}:=\frac{v_{i}}{\sum_{j=1}^{\rel{n}}v_{j}}.  \label{eq:normVote}
\end{equation}
However, while they have established that $(\ref{eq:seatAlloc})$ is accurate as an empirical regularity, they did not explain why it works. In this article, we fill that gap by proving that the seat allocation formula $(\ref{eq:seatAlloc})$ holds under its assumptions, and demonstrating how some of the latter follow from certain more fundamental probabilistic assumptions about vote distribution in district-based elections.

In Sec.~\ref{secJDH} we discuss the Jefferson--D'Hondt method and some of its mathematical properties. In Sec.~\ref{secLit} we summarize the prior work on seat bias. In Sec.~\ref{secProof} we formalize our main result as Theorem \ref{thm:main} and prove it. In Sec.~\ref{secA1} we present a probabilistic model under which some of the technical assumptions appearing in Theorem \ref{thm:main} are asymptotically justified.
Finally, in Sec.~\ref{secTholds} we return to the concept of party relevance, using it to define the natural threshold of representation and derive certain properties thereof. 

Throughout the remainder of this article, 
we use lowercase letters for numbers and density functions, uppercase letters for sets, random variables, and cumulative distribution functions, and bold font for vectors. For variables defined by aggregating over parties, the hat symbol denotes aggregation over relevant parties only, while its absence denotes aggregation over all parties.


\begin{notation}
Let:

\begin{itemize}
\setlength\itemsep{0.2em}
\item $\all{n}\in\mathbb{N}_{+}$ be the \emph{number of parties};

\item $s\in \mathbb{N}_{+}$ be the \emph{number of seats} to be allocated;

\item $c\in\mathbb{N}_{+}$ be the \emph{number of districts};

\item $m_{k}$ be the \emph{magnitude of the $k$-th district}, $k = 1, \dots, c$, i.e., the number of seats to be allocated in that district; note that $\sum_{k=1}^{c} m_{k} = s$;

\item $m:=s/c$ be the \emph{mean district magnitude};


\item $v_{i}$ be the \emph{aggregate number of votes for the $i$-th party}, $i = 1, \dots, \all{n}$; we assume that parties are sorted degressively by the number of votes;

\item $v_{i}^{k}$ be the \emph{number of votes for the $i$-th party in the $k$-th district}; obviously, $v_{i}=\sum_{k=1}^{c} v_{i}^{k}$;

\item $\rel{n}$ be the \emph{number of relevant parties} given by
\begin{equation}
\rel{n}:=\max \left\{l = 1, \dots, \all{n} : \frac{v_{l}}{\sum_{j=1}^{l} v_{j}} > \frac{1}{2m+l}\right\};
\end{equation}

\item $\all{w}_{k}:=\sum_{i=1}^{\all{n}} v_{i}^{k}$ be the \emph{number of votes cast for all parties in the $k$-th district};

\item $\rel{w}_{k}:=\sum_{i=1}^{\rel{n}} v_{i}^{k}$ be the \emph{number of votes cast for relevant parties in the $k$-th district};

\item $\all{v}:=\sum_{i=1}^{\all{n}} v_{i}=\sum_{k=1}^{c} \all{w}_{k}$ be the \emph{aggregate number of votes cast for all parties};

\item $\rel{v}:=\sum_{i=1}^{\rel{n}} v_{i}=\sum_{k=1}^{c} \rel{w}_{k}$ be the \emph{aggregate number of votes cast for relevant parties};

\item $\all{p}_{i}:=v_{i}/\all{v}$ be the \emph{(non-renormalized) aggregate vote share of the $i$-th party};

\item $\rel{p}_{i}:=v_{i}/\rel{v}$ be the \emph{renormalized aggregate vote share of the $i$-th party};

\item $\all{p}_{i}^{k}:=v_{i}^{k}/\all{w}_{k}$ be the \emph{(non-renormalized) vote share of the $i$-th party in the $k$-th district};

\item $\rel{p}_{i}^{k}:=v_{i}^{k}/\rel{w}_{k}$ be the \emph{renormalized vote share of the $i$-th party in the $k$-th district};

\item $s_{i}^{k}$ be the \emph{number of seats of the $i$-th party in the $k$-th district}; note that $\sum_{i=1}^{\all{n}} s_{i}^{k}=m_{k}$;

\item $s_{i}:=\sum_{k=1}^{c} s_{i}^{k}$ be the \emph{total number of seats of the $i$-th party}; note that $\sum_{i=1}^{\all{n}} s_{i}=s$;

\item $q_{i}^{k}:=s_{i}^{k}/m_{k}$ be the \emph{seat share of the $i$-th party in the $k$-th district};

\item $q_{i}:=s_{i}/s$ be the \emph{aggregate seat share of the $i$-th party}.
\end{itemize}
\end{notation}

\begin{notation}
Moreover, let:

\begin{itemize}
\setlength\itemsep{0.2em}

\item $\Delta_{k}$, $k \in \mathbb{N}_{+}$, be a $(k-1)$-dimensional \emph{unit simplex}, i.e.,
\begin{equation}
\Delta_{k} := \left\{ \mathbf{x} \in \mathbb{R}_{\ge 0}^{k} : \sum_{i=1}^{k} x_i = 1 \right\};
\end{equation}

\item $\mathcal{G}_{k}^{l}$, $k, l \in \mathbb{N}_{+}$, be a $(k-1)$-dimensional \emph{$l$-grid simplex}, i.e.,
\begin{equation}
\mathcal{G}_{k}^{l} := \left\{ \mathbf{x} \in \Delta_{k} : l\,\mathbf{x} \in \mathbb{N}^{k} \right\};
\end{equation}

\item $\left\langle x_i \right\rangle_{i=a}^{b}$ denote the \emph{average} of $x_i$ over $i = a, \dots, b$, $a < b \in \mathbb{N}$, i.e.,
\begin{equation}
\langle x_i \rangle_{i=a}^{b} := \frac{1}{b-a+1} \sum_{i=a}^{b} x_i;
\end{equation}

\item $X_{i}^{\downarrow}$ denote the \emph{$i$-th largest element} of a sequence $(X_j)$.
\end{itemize}
\end{notation}


\section{\label{secJDH}The Jefferson--D'Hondt Method}

\subsection{\label{secDivisors}Divisor Methods}

Let $i=1, \dots, \all{n}$. As in this section we focus solely on apportionment as applied to an individual district, aggregate and district-level variables are indistinguishable, wherefore index $k$ will be omitted.

\begin{definition}[Apportionment method]
    An \emph{apportionment method} is a partial function that maps a vote share vector to a seat share vector, $\mathbf{q} : \Delta_{\all{n}} \rightarrow \mathcal{G}_{\all{n}}^{m}$.
\end{definition}

\begin{definition}[Rounding function]
    A \emph{rounding function} is a non-decreasing function $\rho : \mathbb{R} \rightarrow \mathbb{Z}$ such that $|\rho(x) - x| < 1$ for every $x \in \mathbb{R}$.
\end{definition}

\begin{definition}[Divisor method of apportionment]
    An apportionment method $\mathbf{q}$ is called a \emph{divisor method} if and only if there exists a rounding function $\rho$ such that for every $i = 1, \dots, \all{n}$ the seat share of the $i$-th party is given by
    \begin{equation}
        q_{i} = \rho(\all{p}_{i} / \all{\delta}) / m, \label{divSeats}
    \end{equation}
    where the \emph{divisor} $\all{\delta} \in \mathbb{R}_{+}$ is such that
    \begin{equation}
        \sum_{i=1}^{\all{n}} \rho(\all{p}_{i} / \all{\delta}) = m. \label{divSeatSum}
    \end{equation}
\end{definition}

\begin{remark}
    It is often more convenient to use an equivalent form of (\ref{divSeats}):
    \begin{equation}
        q_{i} = \rho(\all{p}_{i} \all{\mu}) / m,  \label{divSeatsAlt}
    \end{equation}%
    where $\all{\mu}:=1/\all{\delta}$ is called the \emph{multiplier}.
\end{remark}

\begin{proposition} \label{prop:divIntervals}
    If the rounding function $\rho$ is right-continuous, the solution set of (\ref{divSeatSum}) is either a half-open interval $(\all{\delta}_{\inf }, \all{\delta}_{\sup }]$ or an empty set. Furthermore, $q_{i}$ does not depend on the choice of $\all{\delta} \in (\all{\delta}_{\inf }, \all{\delta}_{\sup }]$.
\end{proposition}

\begin{proof}
    If $\rho$ is right-continuous, $\all{\delta} \mapsto \rho(\all{p}_{i}/\all{\delta})$ is left-continuous and non-increasing. Thus, the sum of such functions over $i \in \{1, \dots, \all{n}\}$, i.e., $\all{\delta} \mapsto \sum_{i=1}^{\all{n}} \rho(\all{p}_{i}/\all{\delta})$, is also left-continuous and non-increasing. Accordingly, the preimage of every element of its codomain is necessarily either empty or an interval (due to monotonicity) that is right-half-closed (due to left-continuity). We will denote the set of such intervals by $\mathcal{D}$. Since $\mathcal{D}$ is a partition of $\mathbb{R}_{+}$, each $I \in \mathcal{D}$ must also be left-half-open, as desired. 

    Uniqueness of $q_i$ for every divisor interval $I \in \mathcal{D}$ follows from $\rho(\all{p}_{i} / \all{\delta})$ being non-increasing for every $i = 1, \dots, \all{n}$ and every $\all{\delta} \in I$, and $\sum_{i=1}^{\all{n}} \rho(\all{p}_{i} / \all{\delta})$ being constant over $I$. A sum of weakly monotonic functions is constant if and only if all of them are constant. Thus, $q_i$ does not depend on the choice of $\all{\delta} \in I$ for every $i = 1, \dots, \all{n}$.
\end{proof}

\begin{corollary}
    If the rounding function is right-continuous, $\all{\mu} \in \lbrack \all{\mu}_{\inf }, \all{\mu}_{\sup })$, where $\all{\mu}_{\inf }:=1/\all{\delta}_{\sup }$ and $\all{\mu}_{\sup }:=1/\all{\delta}_{\inf }$.
\end{corollary}

\begin{definition}[Rounding thresholds]
    Let $\rho : \mathbb{R} \rightarrow \mathbb{Z}$ be a right-continuous rounding function. For every $k \in \mathbb{Z}$, the preimage of $k$ under $\rho$ is a left-half-closed interval. The sequence of \emph{rounding thresholds} of $\rho$ is a sequence $(\rho_k)_{k \in \mathbb{Z}}$ whose elements are given by $\rho_k := \min \rho^{-1}(k)$.
\end{definition}

\begin{definition}[Electoral quotients] \label{def:elecQuotients}
    Let $\rho$ be a right-continuous rounding function, and let $(\rho_k)_{k \in \mathbb{Z}}$ be the sequence of its rounding thresholds. The $j$-th \emph{electoral quotient} of the $i$-th party, where $j \in \mathbb{N}_{+}$ and $i = 1, \dots, \all{n}$, is given by $\pi_{i,j} := \all{p}_{i}/\rho_j$.
\end{definition}

Note that for every $i = 1, \dots, \all{n}$ the electoral quotients form a decreasing sequence $(\pi_{i,j})_{j \in \mathbb{N}_{+}}$ such that for every $x \in \mathbb{R}_{+}$ there exist only finitely many terms satisfying $\pi_{i,j} > x$. Let $(Q_k)_{k \in \mathbb{N}_{+}}$ be a ``merged'' sequence given by $Q_{(j-1)\all{n} + i} := \pi_{i, j}$ for every $i = 1, \dots, \all{n}$ and every $j \in \mathbb{N}_{+}$. Since the number of parties is finite, for every $x \in \mathbb{R}_{+}$ there exist only finitely many terms of $(Q_k)_{k \in \mathbb{N}_{+}}$ satisfying $Q_k > x$. Accordingly, for every $l \in \mathbb{N}_{+}$ there exists a (not necessarily unique) $l$-th largest element of $(Q_k)_{k \in \mathbb{N}_{+}}$.

\begin{proposition} \label{prop:elecQuotients}
    If $\rho$ is a right-continuous rounding function, then $\all{\delta}_{\inf }=Q_{m+1}^{\downarrow}$ and $\all{\delta}_{\sup }=Q_{m}^{\downarrow}$.
\end{proposition}

\begin{proof}
    By Proposition \ref{prop:divIntervals}, it is sufficient to establish that for every $k \in \mathbb{N}_{+}$ and every $\all{\delta} \in \mathbb{R}_{+}$ we have $\sum_{i=1}^{\all{n}} \rho(\all{p}_{i}/\all{\delta}) < k$ if and only if $\all{\delta} > Q_{k}^{\downarrow}$.
    
    We begin by showing that for every $i = 1, \dots, {\all{n}}$ and every $\all{\delta} \in \mathbb{R}_{+}$ the following equality holds:
    \begin{equation} \label{eq:prop2eq1}
        \rho(\all{p}_{i}/\all{\delta}) = |\{\pi_{i,j}: j \in \mathbb{N} \textrm{ and } \pi_{i,j} \ge \all{\delta}\}|.
    \end{equation}
    Let $k_{i} := \rho(\all{p}_{i} / \all{\delta})$. Then $\all{p}_{i} / \all{\delta} \in \rho^{-1}(k_i)$. It follows that $\all{p}_{i} / \all{\delta} \ge \min \rho^{-1}(k_i) = \rho_{k_i}$, wherefore $\all{p}_{i} / \rho_{k_i} \ge \all{\delta}$. By the same reasoning, $\all{p}_{i} / \rho_{k_i + 1} < \all{\delta}$. Since $\pi_{i,j}$ is decreasing in $j$, it follows that $\pi_{i,j} \ge \all{\delta}$ if and only if $j \in \{1, \dots, k_i\}$.
    
    Then from (\ref{eq:prop2eq1}) and the definition of $(Q_k)_{k \in \mathbb{N}_{+}}$ for every $\all{\delta} \in \mathbb{R}_{+}$ we obtain
    \begin{equation} \label{eq:prop2eq2}
        \sum_{i=1}^{\all{n}} \rho(\all{p}_{i}/\all{\delta}) = \sum_{i=1}^{\all{n}} |\{\pi_{i,j}: j \in \mathbb{N} \textrm{ and } \pi_{i,j} \ge \all{\delta}\}| = |\{l \in \mathbb{N}_{+} : Q_l \ge \all{\delta}\}|.
    \end{equation}
    
    Fix any $k \in \mathbb{N}_{+}$ and any $\all{\delta} \in \mathbb{R}_{+}$. If $\all{\delta} > Q_{k}^{\downarrow}$, then from (\ref{eq:prop2eq2}) it follows that
    \begin{equation} \label{eq:prop2eq3}
       \sum_{i=1}^{\all{n}} \rho(\all{p}_{i}/\all{\delta}) = |\{l \in \mathbb{N}_{+} : Q_l \ge \all{\delta}\}| \le |\{l \in \mathbb{N}_{+} : Q_l > Q_{k}^{\downarrow}\}| < k.
    \end{equation}

    In the other direction, fix any $k \in \mathbb{N}_{+}$ and any $\all{\delta} \in \mathbb{R}_{+}$ such that $\sum_{i=1}^{\all{n}} \rho(\all{p}_{i}/\all{\delta}) < k$. Assume on the contrary that $\all{\delta} \le Q_{k}^{\downarrow}$. From (\ref{eq:prop2eq2}) it then follows that
    \begin{equation} \label{eq:prop2eq4}
       \sum_{i=1}^{\all{n}} \rho(\all{p}_{i}/\all{\delta}) = |\{l \in \mathbb{N}_{+} : Q_l \le \all{\delta}\}| \ge |\{l \in \mathbb{N}_{+} : Q_l \le Q_{k}^{\downarrow}\}| \ge k,
    \end{equation}
    contradicting the assumption that $\sum_{i=1}^{\all{n}} \rho(\all{p}_{i}/\all{\delta}) < k$, and thereby concluding the proof.
\end{proof}

\begin{remark}[Electoral ties]
    We refer to cases where the set of solutions of (\ref{divSeatSum}) is empty as \emph{electoral ties}, although they do not necessarily involve two parties having the same number of votes.
\end{remark}

\begin{example}
    Fix $\all{n} = 2$, $\all{p}_1 = 2/3$, $\all{p}_2 = 1/3$, and $m = 2$, and let $\rho(x) := \floor{x}$. Then for any $\all{\delta} \in (1/3, 2/3]$ we have $s_1 = 1$, $s_2 = 0$, wherefore $\sum_{i=1}^{\all{n}} s_i < m$. However, for $\all{\delta} \in (2/9, 1/3]$ we have $s_1 = 2$, $s_2 = 1$, and thus $\sum_{i=1}^{\all{n}} s_i > m$. Since the left side of (\ref{divSeatSum}) is weakly decreasing in $\all{\delta}$, we need not check any other intervals to establish that there exists no $\all{\delta} \in \mathbb{R}_{+}$ such that $\sum_{i=1}^{\all{n}} s_i = m$. Thus, the election is tied.
\end{example}

\begin{corollary}
    It follows from Proposition \ref{prop:elecQuotients} that a necessary and sufficient condition for an electoral tie to occur is $Q_{m+1}^{\downarrow} = Q_{m}^{\downarrow}$. Accordingly, electoral ties are not possible if vote shares are incommensurable. If $m < \all{w}$ and vote counts are coprime, electoral ties are also not possible.
\end{corollary}

\begin{remark}[Probability of an electoral tie -- continuous models]
    Assume that vote shares are drawn at random from an absolutely continuous probability distribution on $\Delta_{\all{n}}$. Then the set of points for which electoral ties occur is of Lebesgue measure zero. Thus, we can treat such ties are theoretically negligible.
\end{remark}

\begin{remark}[Probability of an electoral tie -- a discrete toy model]
    Assume that $m < \all{w}$ and that vote counts are drawn independently from $\Unif{\{1, \dots, k\}}$, where $k \in \mathbb{N}_{+}$. The probability of an electoral tie occurring for some value of $m$ is bounded from the above by the probability of vote counts of all parties being coprime, which approaches $1 / \zeta(\all{n})$ as $k \rightarrow \infty$, where $\zeta$ is the Riemann zeta function \citep{Nymann72}.
\end{remark}

\begin{remark}[Probability of an electoral tie -- empirical data]
    In a set of $61,416$ Polish local, regional, and national elections held under divisor rules since 1991 (with each electoral district regarded as an individual election), only $11$ electoral ties occurred. For a more general discussion of the frequency of electoral ties, see \citep{MulliganHunter03}. For the reasons stated in this and the above two remarks, we do not concern ourselves with ties in the present article.
\end{remark}

\begin{remark}[Alternative formulations]
    \citet{Janson14} distinguishes three alternative but equivalent formulations of divisor methods besides (\ref{divSeats}) and (\ref{divSeatsAlt}):
    \begin{itemize}
        \item \textbf{iterative formulation}: seats are distributed iteratively, with $k$-th seat being awarded to the party with the highest \emph{comparative vote share}, defined for the $i$-th party as $v_i / \rho_{z_i + 1}$, where $z_i$ is the number of seats already allocated to the $i$-th party;

        \item \textbf{highest-quotients formulation}: the number of seats of the $i$-th party is given by
        \begin{equation}
            s_i = |\{\pi_{i,j} : j \in \mathbb{N}_{+}\} \cap \{ Q_{1}^{\downarrow}, \dots, Q_{m}^{\downarrow} \}|;
        \end{equation}

        \item \textbf{quotient-separation formulation}: the number of seats of the $i$-th party is such that the $L_1$ norm of the seat count vector, $\lVert \mathbf{s} \rVert_{1}$, equals $m$ and $\max_{i = 1, \dots, {\all{n}}} \pi_{i, s_i+1} < \min_{i = 1, \dots, {\all{n}}} \pi_{i, s_i}$.
    \end{itemize}
\end{remark}

\begin{definition}[Linear divisor methods]
    A divisor method is \emph{linear} if and only if it is induced by a right-continuous rounding function $\rho : \mathbb{R} \rightarrow \mathbb{Z}$ such that the rounding thresholds are equidistant, i.e., there exists some $\beta \in [0, 1)$ such that $\rho_k = k + \beta$ for every $k \in \mathbb{Z}$.
\end{definition}

\subsection{\label{secJDHSub}Jefferson--D'Hondt}

The Jefferson--D'Hondt method is the most popular linear divisor method in use in political elections. Accordingly, we will focus on it throughout the remainder of this paper. Nevertheless, the results presented in the following three sections can be easily generalized to other linear divisor methods.

\begin{definition}[Jefferson--D'Hondt method] \label{def:jdh}
    The \emph{Jefferson--D'Hondt method} is a divisor method of apportionment induced by the floor function $\floor{\cdot}$, i.e., a function $\mathbf{q} : \Delta_{\all{n}} \rightarrow \mathcal{G}_{\all{n}}^{m}$ such that
    \begin{equation}
        q_{i} = \floor{\all{p}_{i} / \all{\delta}} / m,  \label{dhSeats}
    \end{equation}
    for every $i = 1, \dots, {\all{n}}$, where $\all{\delta} \in \mathbb{R_{+}}$ (a \emph{Jefferson--D'Hondt divisor}) is such that
    \begin{equation}
        \sum_{i=1}^{\all{n}} \floor{\all{p}_{i}/\all{\delta}} = m.  \label{dhSeatSum}
    \end{equation}
\end{definition}

\begin{remark}
    It is easy to see that the Jefferson--D'Hondt method is a linear divisor method, with the $k$-th rounding threshold given by $\rho_k = k$.
\end{remark}

\begin{remark}
    Under the original Jefferson proposal, the divisor has been fixed, while the number of seats has been allowed to vary \citep{BalinskiYoung78}. The mathematical properties of the method are otherwise unaltered.
\end{remark}

\begin{remark}
The iterative formulation of the Jefferson--D'Hondt method was first proposed in 1888 by \citet{HagenbachBischoff88,HagenbachBischoff05}. It is used in legislative elections in Luxembourg, and it has been used in the United Kingdom for the European Parliament elections before 2019.
\end{remark}

\begin{remark}
\label{divQuotEquiv} The highest-quotients formulation of the Jefferson--D'Hondt method was first introduced by D'Hondt himself \citeyear{DHondt85} in 1885, and is by far the most popular among legislators and political scientists. For instance, all EU countries employing the D'Hondt method for legislative elections -- except Luxembourg -- employ it in their electoral legislation. Note that this formulation closely resembles an earlier proposal by \citet{BurnitzVarrentrapp63}, who called for a modified version of the Borda count, with each elector ranking no more than $m$ candidates, ranks being translated to scores harmonically, and seats being awarded to $m$ candidates with the highest scores. Should the whole electorate be divided into perfectly disciplined partisan voting blocks, and should each block unanimously vote for the same candidates in the same order, the Burnitz--Varrentrapp method would be equivalent to the Jefferson--D'Hondt method.

It is well known that the highest quotient formulation and the ``standard'' formulation (Definition \ref{def:jdh}) of the Jefferson--D'Hondt method are equivalent, i.e., they always generate an identical allocation of seats. For an early proof, see  \citet{Equer11}. To check it quickly, assume $\kappa _{i}$ $( i=1,\dots,\all{n})$ is the number of seats awarded to the $i$-th party under the highest quotient method. Clearly, $\sum_{i=1}^{\all{n}} \kappa_{i}=m$. Recall that $\all{\delta} \in (Q_{m+1}^{\downarrow},Q_{m}^{\downarrow}]$. It follows that $\all{p}_{i}/\kappa_{i} \geq Q_{m}^{\downarrow} \geq \all{\delta} > Q_{m+1}^{\downarrow} \geq \all{p}_{i}/(\kappa_{i}+1)$, and, in consequence, $\kappa_{i}=\floor{\all{p}_{i}/\all{\delta}}$ for every $i=1,\dots,\all{n}$, as desired.
\end{remark}

\citet{Pukelsheim14,Pukelsheim17} defines a simple algorithm for finding a divisor given some fixed \emph{divisor initialization} $\all{\delta} \in \mathbb{R}_{+}$:

\begin{algorithm}[Jump-and-step]
While $\sum_{i=1}^{\all{n}} \floor{\all{p}_{i}/\all{\delta}} \neq m$ do:

\begin{itemize}
\item if $\sum_{i=1}^{\all{n}} \floor{\all{p}_{i}/\all{\delta}} < m$, set
$\all{\delta} \leftarrow \max \{ \all{p}_{i}/(\floor{\all{p}_{i}/\all{\delta}}+1)\} $;

\item if $\sum_{i=1}^{\all{n}} \floor{\all{p}_{i}/\all{\delta}} > m$, set
$\all{\delta} \leftarrow \min \{ \all{p}_{i} / \floor{\all{p}_{i}/\all{\delta}}\} + \varepsilon $,
where $\varepsilon \in (0,1/m)$.
\end{itemize}
\end{algorithm}

\begin{remark} \label{rem:quotas}
Popular divisor initializations for the Jefferson--D'Hondt method include:

\begin{itemize}
\item the \emph{simple quota}, $\all{\delta}_{m}:=1/m$ 
[\citealp{DHondt82}; cf. \citealp{Hare59}];

\item the \emph{Hagenbach-Bischoff quota}, $\all{\delta}_{m}^{HB} := (\floor{\all{w}/(m+1)} +1) / \all{w}$ 
[\citealp{HagenbachBischoff88,HagenbachBischoff05}; cf. \citealp{Droop81}];

\item the \emph{Gfeller-Joachim-Pukelsheim quota}, $\all{\delta}_{m}^{GP}:=(m+\all{n}/2)^{-1}$ \citep{Gfeller90,Joachim17}.
\end{itemize}

\noindent For an in-depth discussion of the origins and attributions of the most popular electoral quotas, see \citet{Dancisin13}.
\end{remark}

\begin{remark} \label{defDevOptMul}
\citet{HappacherPukelsheim96,HappacherPukelsheim00} have established the Gfeller-Joachim-Pukelsheim quota to have the unique property of being asymptotically unbiased under the Jefferson--D'Hondt method as the district magnitude approaches infinity.
\end{remark}

\begin{remark}
The highest-quotients formulation of the D'Hondt method is informally equivalent to the jump-and-step algorithm with divisor initialization $\all{\delta}=\infty $. This is always the least optimal choice, as for every possible vote share vector it requires exactly $m$ iterations to arrive at a correct divisor.
\end{remark}

\begin{remark}
Let $\lambda_{\all{\mu}} := \sum_{i=1}^{\all{n}} \floor{\all{p}_{i} \all{\mu}} - m$ be the \emph{discrepancy} of the seat allocation under multiplier $\all{\mu}$. The distribution of the discrepancies for the Jefferson--D'Hondt method has attracted much scholarly interest. \citet{Happacher01} has provided an analytical formula for the probability distribution of the discrepancy under the assumption that vote shares are drawn from a uniform distribution on the unit simplex. \citet[Thm.~7.5]{Janson13} has established that as the number of seats approaches infinity, the discrepancy distribution approaches the Euler-Frobenius distribution. Finally, \citet{HeinrichEtAl17} have found that the discrepancy distribution can be approximated by applying standard rounding to a sum of uniformly distributed random variables. By the way of illustration, we plot the discrepancy distribution on the 2-dimensional standard unit simplex in Fig. \ref{fig:discrepancies}.
\begin{figure}[tb]\centering\includegraphics
[width=0.78\textwidth]{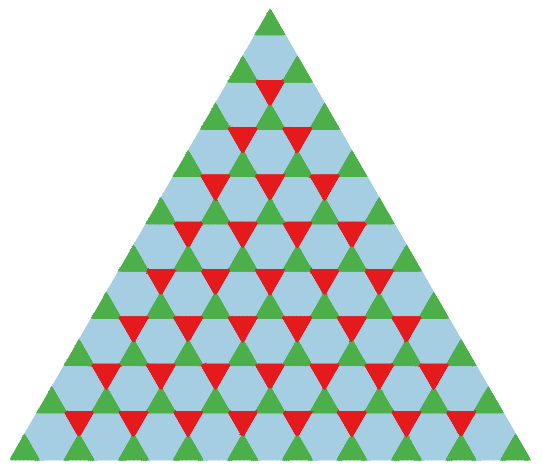}
\caption
{Disrepancy values on the unit simplex $\Delta_{\all{n}}$ for $\all{n}=3$, $m=8$, and $\all{\mu}=m+\all{n}/2=9.5$. Blue, green, and red regions represent, respectively, $\lambda_{\all{\mu}}=0$, $\lambda_{\all{\mu}}=1$, and $\lambda_{\all{\mu}}=-1$.}
\label{fig:discrepancies}
\end{figure}
\end{remark}


\begin{definition}\label{rounding residuals}
    The \emph{rounding residual} of the $i$-th party under multiplier $\all{\mu} \in \lbrack \all{\mu}_{\inf },\all{\mu}_{\sup })$\ is
    \begin{equation}
        \all{r}_{i}(\all{\mu}) := \{ \all{p}_{i} \all{\mu} \} = \all{p}_{i} \all{\mu} - \floor{\all{p}_{i}\all{\mu}} .  \label{residuals}
    \end{equation}
    We can extend this definition for $\all{\mu}_{\sup }$:
    \begin{equation}
        \all{r}_{i} (\all{\mu}_{\sup}) :=\lim_{\all{\mu}\nearrow \all{\mu}_{\sup}} \all{r}_{i}(\all{\mu}).
    \end{equation}
\end{definition}

\begin{lemma}\label{lemma1}
    For every $\all{\mu}\in \lbrack \all{\mu}_{\inf }, \all{\mu}_{\sup })$ and every $k = 1, \dots, \all{n}$ the following equality holds:
    \begin{equation}
        \all{\mu} \sum_{i=1}^{k} \all{p}_{i} = \sum_{i=1}^{k} s_i + \sum_{i=1}^{k} \all{r}_{i}(\all{\mu}).
        \label{mulSumResiduals}
    \end{equation}
\end{lemma}

\begin{proof}
    Recall that by (\ref{divSeatsAlt}) for every $\all{\mu}\in \lbrack \all{\mu}_{\inf },\all{\mu}_{\sup })$ we have $s_{i} = \floor{\all{p}_{i} \all{\mu}}$. Thus, summing (\ref{residuals}) over $i=1,\dots,k$ we obtain $\sum_{i=1}^{k} \all{r}_{i}(\all{\mu}) = \all{\mu} \sum_{i=1}^{k} \all{p}_{i} - \sum_{i=1}^{k} s_{i}$, as desired.
\end{proof}

\begin{corollary}
    For every $\all{\mu}\in \lbrack \all{\mu}_{\inf }, \all{\mu}_{\sup })$ the following equality holds:
    \begin{equation}
        \all{\mu} = m + \sum_{i=1}^{\all{n}} \all{r}_{i}(\all{\mu}).
    \end{equation}
\end{corollary}

\begin{lemma} \label{lemma2}
    For every party $i=1,\dots ,\all{n}$, every $k = 1, \dots, \all{n}$, and every $\all{\mu} \in \lbrack \all{\mu}_{\inf }, \all{\mu}_{\sup })$ the following equality holds:
    \begin{equation}
        s_{i} = \frac{\all{p}_{i}}{\sum_{j=1}^{k} \all{p}_{j}} \left( \sum_{j=1}^{k} s_j + \sum_{j=1}^{k} \all{r}_{j}(\all{\mu})\right) - \all{r}_{i}(\all{\mu}) .  \label{districtEq}
    \end{equation}
\end{lemma}

\begin{proof}
    Note that for every $x \in \mathbb{R}_{+}$, seat allocations and rounding residuals are invariant under simultaneous multiplication of the vote shares by $x$ and division of the multiplier by $x$. Then, (\ref{districtEq}) follows immediately from Lemma \ref{lemma1} and (\ref{residuals}).
\end{proof}

\medskip Recall that in (\ref{eq:relevParties}) we have introduced the concept of \emph{relevant parties}. Note that if the sum of seats over relevant parties, $\sum_{i=1}^{\rel{n}} s_i$, equals $m$, we can express the preceding results in terms of renormalized vote shares, renormalized multipliers, and renormalized rounding residuals:

\begin{definition}[Renormalized multipliers]
    For every multiplier $\all{\mu} \in \mathbb{R}_{+}$ the corresponding \emph{renormalized multiplier} is given by
    \begin{equation}
        \rel{\mu} := \all{\mu} \sum_{i=1}^{\rel{n}} \all{p}_{i}.
    \end{equation}
    In particular,
    \begin{equation}
        \rel{\mu}_{\inf} := \all{\mu}_{\inf} \sum_{i=1}^{\rel{n}} \all{p}_{i}
    \textrm{\quad and \quad}
        \rel{\mu}_{\sup} := \all{\mu}_{\sup} \sum_{i=1}^{\rel{n}} \all{p}_{i}.
    \end{equation}
\end{definition}

\begin{definition}[Renormalized rounding residuals]
    For every party $i = 1, \dots, \rel{n}$ its \emph{renormalized rounding residual} under renormalized multiplier $\rel{\mu}$ is given by
    \begin{equation}
        \rel{r}_i(\rel{\mu}) := \{ \rel{p}_i \rel{\mu} \}.
        \label{eq:renormResid}
    \end{equation}
\end{definition}

\begin{observation}
    Note that for every $i = 1, \dots, \rel{n}$ the following equality holds:
    \begin{equation}
        \rel{p}_{i} = \frac{v_{i}}{\rel{w}} = \frac{\all{w} \all{p}_{i}}{\sum_{j=1}^{\rel{n}} \all{w} \all{p}_{j}} = \frac{\all{p}_{i}}{\sum_{j=1}^{\rel{n}} \all{p}_{j}}.
    \end{equation}
\end{observation}

\begin{observation}
    Note that for every party $i = 1, \dots, \rel{n}$ we have
    \begin{equation}
        \floor{\rel{p}_i \rel{\mu}} = \floor{\all{p}_i \all{\mu}} = s_i
        \label{eq:seatsRelAll}
    \end{equation}
    and 
    \begin{equation} \label{eq:residRelAll}
        \rel{r}_i(\rel{\mu}) = \{ \rel{p}_i \rel{\mu} \} = \{ \all{p}_i \all{\mu} \} = \all{r}_i(\all{\mu}).
    \end{equation}
\end{observation}

\begin{proposition} \label{lemma2rel}
    If $\sum_{j=1}^{\rel{n}} s_j = m$, (\ref{districtEq}) simplifies to:
    \begin{equation}
        s_{i} = \rel{p}_{i} \left( m + \sum_{j=1}^{\rel{n}} \rel{r}_{j}(\rel{\mu})\right) - \rel{r}_{i}(\rel{\mu}). \label{districtEqN}
    \end{equation}
\end{proposition}

\section{\label{secLit}Prior Research on Seat Bias}

It is well known that the Jefferson--D'Hondt method is biased in favor of larger parties 
\citep[see, e.g.,][]{BalinskiYoung01,Benoit00,Carstairs80,Gallagher91,Humphreys11,Huntington21,Huntington28,Huntington31,Lijphart90,MarshallEtAl02,MorseEtAl48,OyamaIchimori95,Pukelsheim14,Rae67,TaageperaLaakso80,TaageperaShugart89,Woodall86,VanEckEtAl05}.
\citet{SainteLague10}
was the first to quantify this effect, finding that the expected seat bias equals $\log 2-1/2$ under the assumptions that $\all{n} = 2$ and $\all{p}_{2}/\all{p}_{1}$ is uniformly distributed over $(0,1)$. 
\citet{Polya18,Polya18a,Polya19,Polya19a,Polya19b} has employed geometric approach to calculate expected seat biases for three-party elections, assuming instead a uniform distribution of the vector of party vote shares over the probability simplex. This line of research has been continued by \citet{SchusterEtAl03}, Drton and Schwingenschlögl \citep{DrtonSchwingenschlogl05,SchwingenschloglDrton04}, and \citet{Schwingenschlogl08}, who have obtained analytical expression for the expected seat bias of the $k$-th largest party in an $\all{n}$-party election under the uniform distribution of the vote share vector\footnote{It should be noted that their results for the expected seat bias of the $k$-th largest party match exactly the results produced by our seat bias formula when $\rel{p}_{i}$ equals the expected vote share of the $k$-th largest party.}.
However, one is frequently interested in estimating the expected seat bias of a specific party (characterized by a given vote share) rather than for an average $k$-th largest party. Moreover, the assumption about the uniform distribution is of uncertain empirical validity.

Analytical formulae for the expected seat bias of the $i$-th party in single-district elections conditioned on that party's vote share have been proposed by \citet{Bochsler10}, \citet{Janson14}, and \citet{Pukelsheim14}. Prima facie, they appear identical to each other and very similar to our seat bias formula. However, despite those similarities, they address different problems and employ different assumptions. Under the assumption that the vote shares follow an arbitrary absolutely continuous distribution over the probability simplex, \citet[Sec.~6.10]{Pukelsheim14} has proven that the rounding residuals converge in distribution to a vector of $\all{n}$ variables drawn from a uniform distribution and stochastically independent of each other and of the party vote shares as the district magnitude approaches infinity\footnote{The proof is a more general case of an earlier proof by \citet{Tukey38}, who has established that the rounding residuals of a scalar variable converge in distribution to a uniform distribution on $(0,1)$. See also  \citet{HeinrichEtAl04}.}. Hence he has deduced \citep[Sec.~7.3]{Pukelsheim14} that the seat biases approach those given by (\ref{seatBias}) in this case. 
\citet[Thm.~3.4]{Janson14} has shown that for any choice of vote shares (under the mild assumption that they are linearly independent over rationals), the expected seat biases for the district magnitude randomly drawn from the uniform distribution on  $\{1,\dots, \eta\}$, where $\eta \in \mathbb{N}_{+}$, also converge to those given by (\ref{seatBias}) as $\eta$ approaches infinity. 
\citet[p.~621]{Bochsler10} has obtained the single-district expected seat bias formula by assuming that the rounding residuals are independent of each party's vote share and always have an expected value of $1/2$.

While all three works discussed in the preceding paragraph significantly advanced our knowledge of seat biases, they share some common limitations. The single-district formulae described by the foregoing authors are only correct in the asymptotic sense -- as the district magnitude approaches infinity\footnote{Technically, Janson treats the district magnitude as randomly drawn from a discrete uniform distribution on $(0,\eta)$, with $\eta$ approaching infinity. But in such case the expected district magnitude also approaches infinity, while the value of its cumulative distribution function at $x$ approaches $0$ for every $x \in \mathbb{R}_{+}$.}, and little is known regarding their respective rates of convergence. 
Moreover, numerical simulations demonstrate that under realistic distributional assumptions, said rates are slow enough as to render them of limited usefulness if the district magnitude is of the order of $3$ to $20$, as is usually the case in real-life elections. It is primarily in this area that we seek to advance prior knowledge by demonstrating that when seat allocations and seat biases are summed over multiple districts, restrictive assumptions about the rounding residuals on which Bochsler, Janson, and Pukelsheim rely can be exchanged for more liberal ones, dealing only with interdistrict averages of residuals (which converge to $1/2$ far more rapidly).

While Theorem \ref{thm:main} is superficially similar to the the results obtained by Janson, Pukelsheim, and Bochsler, it is not merely a generalization of the latter three to the multi-district case. Indeed, the difference is one of kind. First, under assumptions \textbf{A1} to \textbf{A3} formulae (\ref{seatShare}) and (\ref{seatBias}) are deterministic rather than probabilistic. Second, if we relax assumption \textbf{A1a} and switch to the probabilistic model described in Sec.~\ref{secA1}, we still avoid the problematic assumption that $m\rightarrow \infty$, instead providing an error bound for arbitrarily small values of $m\geq 1$.

\section{\label{secProof}The Seat Allocation Theorem}

We have introduced all concepts necessary to state our main result regarding the relationship between the seat share and the renormalized vote share for each relevant party under three assumptions: that there exists such a selection of renormalized multipliers, one for every district, that for every relevant party the rounding residuals average to $1/2$ over all districts (\textbf{A1a}) and the renormalized district vote shares of that party are not correlated with the multipliers (\textbf{A1b}); that all seats are distributed only among relevant parties (\textbf{A2}); and that for every relevant party renormalized district vote shares are not correlated with the numbers of votes for relevant parties (\textbf{A3}).

Throughout this section, we only deal with renormalized vote shares, renormalized multipliers, and renormalized rounding residuals. Since there is no risk of confusion, to avoid excessive verbiage, we omit the term ``renormalized.''

\begin{theorem}
\label{thm:main}Assume that $\rel{n}$ is the number of relevant parties given by (\ref{eq:relevParties}). If:

\begin{enumerate}[topsep=0pt,itemindent=.16cm,labelwidth=\itemindent]
\item[(\textbf{A1})] there exists a sequence $(\rel{\mu}_{k})_{k=1}^{c} \in \prod\limits_{k=1}^{c}\left[\rel{\mu}_{\inf }^{k},\rel{\mu}_{\sup }^{k}\right)$ such that for every party $i=1,\dots,\rel{n}$
\begin{enumerate}[topsep=-6pt,partopsep=0pt,parsep=0pt]
\item[(\textbf{A1a})] \label{A1a} $\left\langle \rel{r}_{i}^{k}\left( \rel{\mu}_{k}\right)\right\rangle_{k=1}^{c} =\frac{1}{2}$;\smallskip
\item[(\textbf{A1b})] \label{A1b} $\mathrm{Cov}\left(\rel{\textbf{p}}_{i},\rel{\boldsymbol{\mu}}\right) =0$, where $\rel{\textbf{p}}_{i} := \left(\rel{p}_{i}^{1}, \dots, \rel{p}_{i}^{c}\right)$ and $\rel{\boldsymbol{\mu}} := \left(\rel{\mu}_{1}, \dots, \rel{\mu}_{c}\right)$;
\end{enumerate}

\item[(\textbf{A2})] $\sum_{i=1}^{\rel{n}} s_{i}^{k}=m_{k}$ for every district $k=1,\dots,c$; and

\item[(\textbf{A3})] $\mathrm{Cov}\left(\rel{\textbf{p}}_{i}, \rel{\textbf{w}}\right) =0$, where $\rel{\textbf{p}}_{i} = \left(\rel{p}_{i}^{1}, \dots, \rel{p}_{i}^{c}\right)$ and $\rel{\textbf{w}} := \left(\rel{w}_{1}, \dots, \rel{w}_{c}\right)$, for every party $i=1,\dots ,\rel{n}$;
\end{enumerate}

\smallskip \noindent then
\begin{equation}
q_{i}=\rel{p}_{i}+\rel{p}_{i}\frac{\rel{n}}{2m}-\frac{1}{2m},  \label{seatShare}
\end{equation}

\smallskip \noindent for every $i=1,\dots ,\rel{n}$,\ where $\rel{p}_{i}$ is the renormalized vote share of the $i$-th party given by (\ref{eq:normVote}).
\end{theorem}

\begin{corollary}
Under assumptions \textbf{A1}-\textbf{A3} the \emph{seat bias} of the $i$-th party, $\rel{\sigma}_{i}:=q_{i}-\rel{p}_{i}$, i.e., the difference between its seat share and its vote share, is given by\footnote{For an alternative form of (\ref{seatBias}), see (\ref{seatBiasAlt}).}
\begin{equation}
\rel{\sigma}_{i}=\frac{\rel{n}}{2m}\left(\rel{p}_{i}-\frac{1}{\rel{n}}\right) .  \label{seatBias}
\end{equation}
\end{corollary}

For the proof of Theorem \ref{thm:main}, we need the following lemma:

\begin{lemma}
\label{lemma3}Under assumptions \textbf{A1}-\textbf{A3} the mean multiplier $\langle\rel{\mu}_{k}\rangle_{k=1}^{c}=m+\rel{n}/2$, i.e., the inverse of the Gfeller-Joachim-Pukelsheim quota (Remark \ref{rem:quotas}) with non-relevant parties excluded.
\end{lemma}

\begin{proof}
By \textbf{A2}, $\sum_{i=1}^{\rel{n}} s_{i}^{k} = m_{k}$. Hence, from Lemma \ref{lemma1} and (\ref{eq:residRelAll}) we obtain $\sum_{i=1}^{\rel{n}} \rel{r}_{i}^{k}(\rel{\mu}_{k})=\rel{\mu}_{k}-m_{k}$. Summing this over all districts, we arrive at
\begin{equation}
\sum_{k=1}^{c} \sum_{i=1}^{\rel{n}} \rel{r}_{i}^{k}(\rel{\mu}_{k}) =
\sum_{k=1}^{c} \left(\rel{\mu}_{k}-m_{k}\right) = 
c\langle\rel{\mu}_{k}\rangle_{k=1}^{c} - s.
\end{equation}
On the other hand, from \textbf{A1a} it follows that $\sum_{k=1}^{c} \rel{r}_{i}^{k}(\rel{\mu}_{k})=c/2$ for every $i=1,\dots ,\rel{n}$. Therefore,
\begin{equation}
\sum_{i=1}^{\rel{n}} \sum_{k=1}^{c} \rel{r}_{i}^{k}(\rel{\mu}_{k}) = 
\sum_{i=1}^{\rel{n}} \frac{c}{2} = \frac{c \rel{n}}{2}.
\end{equation}
Accordingly,
\begin{equation}
\langle\rel{\mu}_{k}\rangle_{k=1}^{c} =m + \frac{\rel{n}}{2},  \label{meanDiv}
\end{equation}
as desired.
\end{proof}

We can now proceed to the proof of Theorem \ref{thm:main}.

\begin{proof}
Fix any $i=1,\dots ,\rel{n}$. Note that by (\ref{eq:renormResid}) and (\ref{eq:seatsRelAll}),
\begin{equation}
s_{i}^{k} = \rel{p}_{i}^{k} \rel{\mu}_{k} - \rel{r}_{i}^{k}(\rel{\mu}_{k}).
\label{districtEq2}
\end{equation}
Taking a sum over all districts, we obtain
\begin{equation}
s_{i}=\sum_{k=1}^{c} \Big(\rel{p}_{i}^{k} \rel{\mu}_{k} - \rel{r}_{i}^{k}(\rel{\mu}_{k})\Big) =
\sum_{k=1}^{c} \rel{p}_{i}^{k} \rel{\mu}_{k} - \sum_{k=1}^{c} \rel{r}_{i}^{k}(\rel{\mu}_{k}).
\label{aggrDistrictEq}
\end{equation}
From \textbf{A1b} and Lemma \ref{lemma3} it follows that
\begin{equation}
\sum_{k=1}^{c} \rel{p}_{i}^{k} \rel{\mu}_{k} = c \langle \rel{p}_{i}^{k}\rangle_{k=1}^{c} \langle\rel{\mu}_{k}\rangle_{k=1}^{c} = \langle \rel{p}_{i}^{k}\rangle_{k=1}^{c} \left(s + \frac{c\rel{n}}{2}\right).
\end{equation}
However, by \textbf{A3},
\begin{equation}
\langle \rel{p}_{i}^{k}\rangle_{k=1}^{c} = \frac{\sum_{k=1}^{c} v_i^k}{\sum_{k=1}^{c}{\rel{w}_k}} = \frac{v_i}{\rel{v}} = \rel{p}_{i}.
\end{equation}
Thus,
\begin{equation}
s_{i}=\rel{p}_{i}\left( s+\frac{c\rel{n}}{2}\right) -\sum_{k=1}^{c} \rel{r}_{i}^{k}(\rel{\mu}_k). \label{eq:410}
\end{equation}
Finally, by \textbf{A1a}, we have $\sum_{k=1}^{c} \rel{r}_{i}^{k}(\rel{\mu}_k)=c/2.$ Substituting this into (\ref{eq:410}) and dividing both sides by $s$, we thus arrive at the seat allocation formula (\ref{seatShare}):
\begin{equation}
q_{i}=\rel{p}_{i}\left( 1+\frac{\rel{n}}{2m}\right) -\frac{1}{2m}=\rel{p}_{i} + \rel{p}_{i}\frac{\rel{n}}{2m} -\frac{1}{2m}.
\end{equation}
\end{proof}

\begin{observation}
    As an immediate consequence of \textbf{A2}, if $\rel{n} < \all{n}$, then
    \begin{equation}
        q_i = 0
    \end{equation}
    for every non-relevant party $i = \rel{n} + 1, \dots, \all{n}$.
    \label{obs:nonRelevant}
\end{observation}

\section{\label{secA1}Discussion of Assumption A1}

Assumptions \textbf{A2} and \textbf{A3} are essentially only for the convenience of application of Theorem \ref{thm:main}, as it can be expressed in terms of the average renormalized district-level vote share of the $i$-th party, $\left\langle \rel{p}_{i}^{k}\right\rangle_{k=1}^{c}$, and the sum of seat shares of the non-relevant parties, $\sum_{j=\rel{n}+1}^{\all{n}} q_{j}$. Assumptions \textbf{A2} and \textbf{A3} would then be superfluous, as we would obtain the following corollary:

\begin{corollary}
    If \textbf{A1}, then
    \begin{equation}
        q_{i} = \left\langle \rel{p}_{i}^{k}\right\rangle_{k=1}^{c} \left(1 - \sum_{j=\rel{n}+1}^{\all{n}} q_{j} + \frac{\rel{n}}{2m}\right) - \frac{1}{2m},
    \end{equation}
    for every $i = 1, \dots, \rel{n}$.
    \label{cor:a1only}
\end{corollary}

The principal advantage of Theorem \ref{thm:main} over Corollary \ref{cor:a1only} is purely practical: in the context of most expected applications, $\rel{p}_{i}$ is known, while $\left\langle \rel{p}_{i}^{k}\right\rangle_{k=1}^{c}$ and $\sum_{j=\rel{n}+1}^{\all{n}} q_{j}$ are not. However, the above makes it clear that \textbf{A2} and \textbf{A3} do not give significant insight into the workings of the Jefferson--D'Hondt method.

On the other hand, assumption \textbf{A1} is of fundamental importance to Theorem \ref{thm:main} and cannot be avoided in a manner similar to the other ones. At the same time, it does not easily correspond to normative intuitions about electoral systems, instead requiring some additional justification. In this section, we seek to partially provide such a justification by demonstrating that, under a probabilistic model of elections involving certain reasonable distributional assumptions, we can expect at least \textbf{A1a} to be approximately satisfied if the number of districts is sufficiently large.

As a consequence of Lemma \ref{lemma2} and \textbf{A2}, non-relevant parties have no effect on seat allocation among relevant parties under Theorem \ref{thm:main}. It follows that they can be disregarded at will. Accordingly, a probabilistic model of relevant parties only is sufficient for our present purposes. It follows that in this section we need not distinguish between renormalized and non-renormalized variables.

Let us assume that an election conforms to the following probabilistic model:

\begin{itemize}
\item let $c\in \mathbb{N}_{+}$ be the number of districts;

\item let district magnitudes $M_{1},\dots ,M_{c}$ be independent random variables identically distributed according to some discrete probability distribution $\mathcal{M}$ on $\mathbb{N}_{+}$ with an expectation $m \in [m_{\min},\infty)$, where $m_{\min}\geq 1$;


\item let $\mathbf{p}:=\left( p_{1},\dots ,p_{n}\right) \in \Delta _{n}$ be a vector of aggregate vote shares such that $p_{i} > \left( 2m+n\right) ^{-1}$ for every $i=1,\dots ,n$, i.e., that all parties are relevant;

\item let $\mathcal{V}$ be an absolutely continuous probability distribution on $\Delta_{n}$ with an expectation $\mathbf{p}$ and a continuously differentiable density $f_{\mathcal{V}}$ vanishing at the faces of $\Delta_{n}$;


\item let district-level vote share vectors $\mathbf{P}_{1},\dots ,\mathbf{P}_{c}$ be independent random variables identically distributed according to $\mathcal{V}$.
\end{itemize}

Assume further that $M_{k}$ and $\mathbf{P}_{k}$ are independent for every $k=1,\dots ,c$. \smallskip

Among the assumptions of this model, the one regarding the density of $\mathcal{V}$ appears most in need of an additional justification. It comes from an empirical observation that in most elections, a relevant party almost never comes close to winning all (or no) votes in any district. Hence, the assumption that $f_{\mathcal{V}}$ vanishes at the faces of $\Delta_n$ is consistent with real-life voting patterns.

Note that by the strong law of large numbers, the average district magnitude $\left\langle M_{k}\right\rangle_{k=1}^{c} \overset{\mathrm{a.s.}}{\rightarrow }m$, and the average vote share vector $\left\langle \mathbf{P}_{k}\right\rangle_{k=1}^{c} \overset{\mathrm{a.s.}}{\rightarrow }\mathbf{p}$ as $c\rightarrow \infty $.

Let $\Theta$ be an interval-valued function of $(a,\mathbf{b}) \in \mathbb{N}_{+} \times \Delta_{n}$, mapping a district magnitude $a$ and a vote share vector $\mathbf{b} := (b_1, \dots, b_n)$ to a multiplier interval $[1/Q_{a}^{\downarrow}, 1/Q_{a+1}^{\downarrow})$, where $(Q_{n})_{n \in \mathbb{N}_{+}}$ is given by $Q_{(j-1)n+i}:=b_{i}/j$ for every $i = 1, \dots, n$ and $j \in \mathbb{N}_{+}$. Let $\theta$ be a selection of $\Theta$, i.e., a function $\mathbb{N}_{+} \times \Delta_{n} \rightarrow \mathbb{R}_{+}$ such that $\theta (a,\mathbf{b}) \in \Theta (a,\mathbf{b})$ for every $(a,\mathbf{b}) \in \mathbb{N}_{+} \times \Delta_{n}$. Finally, within each district $k=1,\dots ,c$, let $U_{k}$ be a random variable given by $U_{k} := \theta( M_{k},\mathbf{P}_{k})$. For \textbf{A1} to hold, we need to demonstrate that there exists such $\theta$ that $\mathbb{E}\{ P_{i}^{k}U_{k}\} =1/2$ and $\mathrm{Cov}(P_{i}^{k},U_{k}) =0$ for every $i=1,\dots ,n$ and every $k = 1, \dots, c$. This we are unable to do without further assumptions as to $\mathcal{V}$. However, using Theorem \ref{thm:a1} below we demonstrate that regardless of the choice of $\theta$, assumption \textbf{A1a} is satisfied approximately in the limit of $c\rightarrow \infty $. We leave the question of demonstrating that \textbf{A1b} is also satisfied approximately for future work, although we note that the chequered pattern of the discrepancies on the probability simplex (as illustrated for $n=3$ by Fig. \ref{fig:discrepancies} above) reveals one promising avenue of approach.

\begin{theorem}
\label{thm:a1}If:

\begin{enumerate}[topsep=0pt,itemindent=.16cm,labelwidth=\itemindent]
\item[(\textbf{B1})] $P$ is a random variable with an absolutely continuous distribution supported on a subset of $\left[ 0,1\right] $ with a continuously differentiable density $f_{P}$;

\item[(\textbf{B2})] $U$ is a random variable with a mixed discrete-continuous distribution supported on a subset of $[u,\infty) $, where $u \in \mathbb{R}_{+}$, with a cumulative distribution function $\Psi$, density $\psi$, and a finite set of probability atoms $A$, such that $\psi(a)=0$ for every $a \in A$; and

\item[(\textbf{B3})] $f_P(0) = 0$ and $f_P(1) = 0$;
\end{enumerate}

\noindent then
\begin{equation}
\left\vert \mathbb{E}(\{PU\}) -\frac{1}{2}\right\vert
\leq \frac{1}{12u} V(P), \label{a5bound}
\end{equation}
\noindent where $V(P) :=\int_{0}^{1} \vert f_{P}^{\prime}(x)\vert \,dx$ is the total variation of $f_{P}$.
\end{theorem}

\begin{proof}
Let $f_{P|U}$ be the conditional density of $P$ with respect to $U$ and let $F_{P|U}$ be the conditional cumulative distribution function of $P$ with respect to $U$. Let $R:= \{PU\}$ and consider its conditional distribution with respect to $U$. Clearly,
\begin{equation}
\Pr (R \le x | U = y) = \sum_{l=0}^{\floor{y}} \Pr (PU \in [l, l+x] \,|\, U = y).
\end{equation}
Thus, the conditional density of $R$ with respect to $U$, $\varphi_{R|U} : [0,1) \rightarrow \mathbb{R}_{\ge 0}$, is given by
\begin{equation}
\varphi_{R|U} (x|y) := \frac{1}{y} \sum_{l=0}^{\floor{y}} f_{P|U}\bigg(\frac{l+x}{y} \bigg|\, y\bigg).
\label{densResidCond}
\end{equation}
Accordingly, the unconditional density of $R$ is given by $\varphi: [0, 1) \rightarrow \mathbb{R}_{\ge 0}$ defined as
\begin{equation}
\varphi_{R} (x) :=\int_{u}^{\infty}\frac{1}{y}\sum_{l=0}^{\floor{y}} f_{P|U} \bigg(\frac{l+x}{y} \bigg|\, y \bigg) \,d\Psi \left(y\right).  \label{densResid}
\end{equation}
In consequence, the expected value of $R$ equals
\begin{equation}
\mathbb{E}\left( R\right) =\int_{0}^{1}\int_{u}^{\infty }\frac{x}{y} \sum_{l=0}^{\floor{y}} f_{P|U}\bigg(\frac{l+x}{y} \bigg|\, y \bigg) \,d\Psi \left( y\right) \,dx.  \label{expectResid}
\end{equation}%
Let us substitute $z:=(l+x)/y$ and change the order of integration and summation in (\ref{expectResid}) to obtain
\begin{align}
\mathbb{E}(R) & =
\int_{u}^{\infty} \sum_{l=0}^{\floor{y}} \int_{\frac{l}{y}}^{\frac{l+1}{y}}
\frac{yz-l}{y} f_{P|U}(z|y) \,y\,dz\,d\Psi \left( y\right) \notag \\
& =\int_{u}^{\infty }\int_{0}^{\frac{\floor{y} +1}{y}}yz\,f_{P|U}(z|y) \,dz\,d\Psi (y) -
\int_{u}^{\infty} \sum_{l=0}^{\floor{y}} \int_{\frac{l}{y}}^{\frac{l+1}{y}} l\,f_{P|U} (z|y) \,dz\,d\Psi (y) \notag \\
& =\mathbb{E}(PU) -\int_{u}^{\infty} \sum_{l=0}^{\floor{y}} l \left(F_{P|U}\Big(\frac{l+1}{y} \Big|\, y\Big) -F_{P|U}\Big(\frac{l}{y} \Big|\, y\Big)\right) \,d\Psi (y) \notag \\
& =\mathbb{E}(PU) -\int_{u}^{\infty } \bigg(\floor{y} F_{P|U}\Big( \frac{\floor{y} +1}{y} \Big|\, y\Big) -\sum_{l=0}^{\floor{y}}F_{P|U} \Big(\frac{l}{y} \Big|\, y\Big) \bigg) \,d\Psi (y) \notag \\
& =\mathbb{E}(PU) -\mathbb{E}(\floor{U}) +\int_{u}^{\infty }\sum_{l=0}^{\floor{y}} F_{P|U}\bigg(\frac{l}{y} \bigg|\, y\bigg) \,d\Psi (y) \notag \\
& =\mathbb{E}(PU) -\mathbb{E}(\floor{U}) - 1 +\int_{u}^{\infty }\sum_{l=0}^{\floor{y}+1} F_{P|U}\bigg(\frac{l}{y} \bigg|\, y\bigg) \,d\Psi (y).
\label{expectResid2}
\end{align}

\noindent Let $\Phi_{P|U}(l|y) :=F_{P|U}(l/y|y)$ for every $0 \leq l\leq \floor{y}+1$. If $\Phi_{P|U}$ is smooth of class $C^{2h}$, $h\in \mathbb{N}_{+}$, from the Euler-Maclaurin summation formula we obtain
\begin{align}
\sum_{l=0}^{\floor{y}+1} \Phi_{P|U}(l|y) &=
\int_{0}^{\floor{y}+1} \Phi_{P|U}(x|y) \,dx +
\frac{\Phi_{P|U}(\floor{y}+1|\,y) - \Phi_{P|U}(0|y) }{2} \notag \\
& +\sum_{j=1}^{h}\frac{B_{2j}}{(2j)!}\left(\Phi_{P|U}^{(2j-1)} (\floor{y}+1|\,y) - \Phi_{P|U}^{(2j-1)}(0|y) \right) + \omega_{y}^{2h}, \label{eulerMaclaurin}
\end{align}
where, for every $k=2,\dots ,h$, $B_{k}$ is the $k$-th \emph{Bernoulli number}, $B_{k}(x)$ is the $k$-th \emph{Bernoulli polynomial}, $\Phi_{P|U}^{(k)}$ is the $k$-th derivative of $\Phi_{P|U}$, and $\omega_{y}^{2h}$ is an error term given by
\begin{equation}
\omega_{y}^{2h}:=-\frac{1}{(2h)!} \int_{0}^{\floor{y}+1} \Phi_{P|U}^{(2h)}(x|y) B_{2h} \left(\{x\} \right) \,dx
\end{equation}
\citep[Corollary 9.2.3 (2)]{Cohen07}. Note that
\begin{equation}
\left\vert \omega_{y}^{2h}\right\vert \leq \frac{1}{(2h)!}\left\vert \underset{x \in (0,1)}\max B_{2h}(\{x\}) \right\vert \int_{0}^{\floor{y}+1}\left\vert \Phi_{P|U}^{(2h)}(x|y) \right\vert \,dx.
\end{equation}
\citet[Thm.~1]{Lehmer40} has shown that
\begin{equation}
\left\vert \underset{x\in \left( 0,1\right) }{\max }B_{2h}(\{
x\}) \right\vert \leq \left\vert B_{2h}\right\vert =\frac{%
(2h)!}{(2\pi)^{2h}} 2\zeta (2h) ,
\end{equation}%
where $\zeta $ is the Riemann zeta function. Accordingly,%
\begin{equation}
\left\vert \omega_{y}^{2h}\right\vert \leq \frac{2\zeta(2h) }{%
(2\pi)^{2h}}\int_{0}^{\floor{y}+1}\left\vert
\Phi_{P|U}^{(2h)}(x|y) \right\vert \,dx.
\label{emLastTerm}
\end{equation}%
Substituting $F_{P|U}(l/y|y) $ for $\Phi_{P|U}(l|y)$ in successive terms of (\ref{eulerMaclaurin}) and in (\ref{emLastTerm}), we get:
\begin{eqnarray}
\int_{0}^{\floor{y}+1} \Phi_{P|U}(x|y) \,dx &=&
\int_{0}^{\floor{y}+1} F_{P|U}\Big(\frac{x}{y}\Big|\,y\Big) \,dx =
y\int_{0}^{\frac{\floor{y}+1}{y}} F_{P|U}(z|y) \,dz \notag \\
&=& y\int_{0}^{1} F_{P|U}(z|y) \,dz + y \int_{1}^{\frac{\floor{y}+1}{y}} F_{P|U}(z|y) \,dz \notag \\
&=& y \big(1-\mathbb{E}(P|U=y)\big) + (\floor{y}+1 - y), \label{em1stTerm}
\end{eqnarray}\smallskip
\begin{equation}
\frac{\Phi_{P|U}(\floor{y}+1|\,y) - \Phi_{P|U}(0|y)}{2}=\frac{1}{2} F_{P|U}\bigg(\frac{\floor{y}+1}{y}\bigg|\,y\bigg) = \frac{1}{2},  \label{em2dTerm}
\end{equation}\smallskip
\begin{equation}
\Phi_{P|U}^{(j)}(x|y) =y^{-j} F_{P|U}^{(j)} \Big(\frac{x}{y}\Big|\,y\Big) \text{ for every } j=1,\dots ,2h.
\label{em3dTerm}
\end{equation}%
Therefore,
\begin{align}
\sum_{l=0}^{\floor{y}+1} F_{P|U}\bigg(\frac{l}{y}\bigg|\,y\bigg) &= 
\floor{y} - y \mathbb{E}(P|U=y) + \frac{3}{2} \notag \\
&+\sum_{j=1}^{h}\frac{B_{2j}y^{-(2j-1) }}{\left( 2j\right) !}%
\left(F_{P|U}^{(2j-1)}\bigg(\frac{\floor{y}+1}{y}\bigg|\,y\bigg) - F_{P|U}^{(2j-1)}(0|y) \right) + \omega_{y}^{2h}, \label{eulerMaclaurin3}
\end{align}
where%
\begin{equation}
\left\vert \omega_{y}^{2h}\right\vert \leq \frac{2y\zeta(2h) }{%
(2\pi y) ^{2h}}\int_{0}^{\frac{\floor{y}+1}{y}}\left\vert F_{P|U}^{(2h)}(z|y) \right\vert \,dz.  \label{em4thTerm}
\end{equation}\smallskip

By \textbf{B3} for $h = 1$ we obtain
\begin{gather}
\left. \sum_{j=1}^{h} \frac{B_{2j}y^{-(2j-1)}}{(2j)!} \left(F_{P|U}^{(2j-1)}\bigg( \frac{\floor{y}+1}{y}\bigg|\,y\bigg) - F_{P|U}^{(2j-1)}(0|y) \right) \right\vert _{h=1} \notag \\
=\frac{B_{2}}{2y}\left(f_{P|U}\bigg( \frac{\floor{y}+1}{y}\bigg|\,y\bigg) -f_{P|U}(0|y)\right) = 0.
\end{gather}
Therefore,
\begin{equation}
\left. \sum_{l=0}^{\floor{y}+1} F_{P|U}\bigg(\frac{l}{y}\bigg|\,y\bigg) \right\vert_{h=1} =
\floor{y} - y\mathbb{E}(P|U=y) + \frac{3}{2} + \omega_{y}^{2},
\end{equation}%
where%
\begin{equation}
\left\vert \omega_{y}^{2}\right\vert \leq \frac{2y\zeta(2)}{(2\pi y)^{2}} \int_{0}^{\frac{\floor{y}+1}{y}} \left\vert F_{P|U}^{(2)}(z|y) \right\vert \,dz =
\frac{1}{12y}\int_{0}^{1}\left\vert f_{P|U}^{\prime}(z|y) \right\vert \,dz.
\end{equation}

By incorporating the foregoing result into (\ref{expectResid2}) we obtain
\begin{align}
\mathbb{E}(R) &= \mathbb{E}(PU) -\mathbb{E}(\floor{U}) - 1 + \int_{u}^{\infty} \left(\floor{y} - y\mathbb{E}(P|U=y) + \frac{3}{2} + \omega_{y}^{2}\right) \,d\Psi(y) \notag \\
&= \mathbb{E}(PU) -\mathbb{E}(\floor{U}) - 1 + \mathbb{E}(\floor{U}) -\mathbb{E}(PU) + \frac{3}{2} + \mathbb{E}(\omega_{U}^{2}) 
=\frac{1}{2} + \mathbb{E}(\omega_{U}^{2}).
\label{a5errTerm}
\end{align}
Thus we arrive at
\begin{align}
\left\vert \mathbb{E}(R) -\frac{1}{2}\right\vert \leq \mathbb{E}\bigg(\frac{1}{12U} \int_{0}^{1}\left\vert f_{P|U}^{\prime}(z|U) \right\vert \,dz\bigg) \text{,}  \label{a5bound2}
\end{align}%
where $\int_{0}^{1} |f_{P|U}^{\prime}(z|U)|\,dz$ is the total variation of $f_{P|U}$. Since $U\geq u$, it follows that%
\begin{equation}
\left\vert \mathbb{E}\left( R\right) -\frac{1}{2}\right\vert \leq \frac{1}{12u}\mathbb{E}\bigg(\int_{0}^{1}\left\vert f_{P|U}^{\prime}(z|U) \right\vert \,dz \bigg).
\end{equation}%

For $y\in \mathrm{supp}\, \psi $, we have%
\begin{equation}
f_{P|U}^{\prime}(z|y) =\frac{\partial}{\partial{z}} \frac{f_{P,U}(z,y)}{\psi(y)} =
\frac{1}{\psi(y)} \frac{\partial }{\partial z} f_{P,U}(z,y),
\end{equation}%
where $f_{P,U}$ is the joint density of $P$ and $U$, and
\begin{align}
\int_{\mathrm{supp}\, \psi}\int_{0}^{1}\left\vert f_{P|U}^{\prime} (z|y) \right\vert \,dz\,d\Psi(y) &= \int_{u}^{\infty }\frac{\psi(y)}{\psi(y)}%
\int_{0}^{1}\left\vert \frac{\partial }{\partial z}f_{P,U}(z,y)
\right\vert \,dz\,dy \notag \\
&=\int_{0}^{1}\int_{u}^{\infty }\left\vert \frac{\partial }{\partial z}%
f_{P,U}(z,y) \right\vert \,dy\,dz \notag \\
&=\int_{0}^{1}\left\vert f_{P}^{\prime} (z)
\right\vert \,dz=V(P) .
\end{align}%

On the other hand, for $y\in A$, we have%
\begin{equation}
f_{P|U}^{\prime} (z|y) = \frac{\partial}{\partial{z}} \frac{f_{P,U}(z,y)}{\Pr(U=y)} =
\frac{1}{\Pr(U=y)}\frac{\partial}{\partial z} f_{P,U}(z,y),
\end{equation}%
and%
\begin{align}
\int_{A} \int_{0}^{1}\left\vert f_{P|U}^{\prime} (z|y) \right\vert \,dz\,d\Psi(y) &=
\sum_{y\in A}\frac{\Pr \left( U=y\right) }{\Pr \left( U=y\right) }%
\int_{0}^{1}\left\vert \frac{\partial }{\partial z}f_{P,U}(z,y)
\right\vert \,dz \notag \\
&=\int_{0}^{1}\sum_{y\in A}\left\vert \frac{\partial }{\partial z}%
f_{P,U} (z,y) \right\vert \,dz \notag \\
&=\int_{0}^{1}\left\vert f_{P}^{\prime}(z)
\right\vert \,dz=V(P) .
\end{align}

Accordingly,
\begin{equation}
\mathbb{E}\bigg(\int_{0}^{1}\left\vert f_{P|U}^{\prime}(z|U) \right\vert \,dz  \bigg) = V(P),
\end{equation}
and thus
\begin{equation}
\left\vert \mathbb{E}(R) -\frac{1}{2}\right\vert \leq \frac{1}{12u} V(P),
\end{equation}
as desired.
\end{proof}\medskip


Under the probabilistic model of elections described above, let $P_{i}^{k}$, 
$k=1,\dots ,c$, be the $i$-th barycentric coordinate of $\mathbf{P}_{k}$ (i.e., the vote share of the $i$-th party).
Since $\mathbf{P}_{k}$ has a continuously differentiable density vanishing at the faces of $\Delta _{n}$, it follows that for every $i=1,\dots ,n$ the density of $P_{i}^{k}$ is also continuously differentiable and vanishes at $0$ and $1$.
Thus, we can equate $P_{i}^{k}$ with $P$ in Theorem \ref{thm:a1}, as it satisfies \textbf{B1} and \textbf{B3}. To satisfy \textbf{B2}, fix an arbitrarily small $\varepsilon > 0$ and a selection $\theta$ given by:
\begin{equation}
\theta(a, \mathbf{b}) :=
\begin{cases}
a+n/2 & \textrm{ for } a+n/2 \in \Theta(a, \mathbf{b}), \\ 
\inf \Theta(a, \mathbf{b}) & \textrm{ for } a+n/2 < \inf \Theta(a, \mathbf{b}), \\ 
\sup \Theta(a, \mathbf{b}) - \varepsilon & \textrm{ for } \sup \Theta(a, \mathbf{b}) \leq a+n/2.
\end{cases}
\end{equation}%
Note that $U_{k}$ defined as $\theta(M_{k}, \mathbf{P}_{k})$ is absolutely continuous over $[M_{k}, \infty) \setminus \{M_{k}+n/2\}$, and has a single probability atom at $M_{k}+n/2$, thus satisfying \textbf{B2.}

It follows from Theorem \ref{thm:a1} that $\left\vert \mathbb{E}%
(\{P_{i}^{k} U_{k}\}) -\frac{1}{2}\right\vert $ is bounded from the above
by $\frac{1}{12} M_{k}^{-1} V(P_i^k).$ Since $\mathbf{P}_{1}, \dots, \mathbf{P}_{c}$ are identically distributed, and so are $M_1, \dots, M_c$, from the strong law of large numbers it then follows that
\begin{equation}
\left\langle\left\vert \{P_{i}^{k} U_{k}\} - \frac{1}{2} \right\vert\right\rangle_{k=1}^{c} \overset{\mathrm{a.s.}}{%
\rightarrow} \left\vert \mathbb{E}(\{P_{i}^{l} U_{l}\}) - \frac{1}{2} \right\vert \le \frac{1}{12}\mathbb{E}(M_l^{-1}) V(P_i^l)
\end{equation}
for every $l = 1, \dots, c$.

\begin{remark}
    If the distribution of $\mathbf{P}_{l}$ is unimodal, the total variation of $P_i^l$ equals twice the value of its marginal density, $f_{P_i^l}(x)$, at its mode, i.e., $V(P_i^l) = 2 f_{P_i^l}(\max_{x \in (0, 1)} f_{P_i^l}(x))$ for every $i = 1, \dots, n$.
\end{remark}

For vote distributions encountered in real-life elections, $V(P_i^l)$ rarely exceeds $3$. Meanwhile, for typical district magnitudes, $\mathbb{E}(M_{l}^{-1}) \le 1/5$. Thus, $\frac{1}{12} \mathbb{E}(M_{l}^{-1}) V(P_i^l) \le 1/20$, ensuring that \textbf{A1a} is satisfied approximately in the limit of $c\rightarrow \infty $.

\section{\label{secTholds}Relevant Parties and Natural Thresholds}

Note that formulae (\ref{seatShare}) and (\ref{seatBias}) can only be applied to relevant parties. 
This restriction is connected to another important consequence of the Jefferson--D'Hondt method, namely the existence of a \emph{threshold of representation}, i.e., such $\all{\tau} \in (0, 1)$ that $\all{p}_{i} < \all{\tau}$ implies $\all{q}_{i}=0$. 
To distinguish it from statutory thresholds which are present in some electoral systems, and which operate independently of the Jefferson--D'Hondt method, threshold $\all{\tau}$ is frequently referred to as the \emph{natural threshold}. 

At a district level, the natural threshold follows from Definition \ref{def:jdh}. To see how, recall that by (\ref{dhSeats}), $s_{i}^{k} = \floor{\all{p}_{i}^{k} \all{\mu}_{k}}$. Hence, it is evident that if $\all{p}_{i}^{k} \le 1 / \all{\mu}_{\sup}^{k}$, then $s_{i}^{k}$, and thus also $q_{i}^{k}$, must necessarily be $0$. But $1 / \all{\mu}_{\sup}^{k} = (Q^k)_{m+1}^{\downarrow}$ depends on the vector of vote shares. It would be useful to have an estimate of the natural threshold that depends only on $\all{n}$ and $m_k$. Several authors \citep[see, e.g.,][]{DHondt83,LijphartGibberd77,PalomaresRamirezGonzalez03,RaeEtAl71,Rokkan68} have provided estimates of the lower and upper bounds of the interval in which the natural threshold must fall, known respectively as the \emph{threshold of inclusion} $\tau_{k}^{-} := (\all{n} + m_{k} - 1)^{-1}$ and the \emph{threshold of exclusion} $\tau_{k}^{+} := (m_{k} + 1)^{-1}$. Those can be used to estimate the \emph{aggregate thresholds of exclusion / inclusion}, viz.
\begin{equation}
\tau^{+} := \sum_{k=1}^{c} \frac{1}{m_{k}+1} \frac{\all{v}_{k}}{\all{v}},
\end{equation}%
and
\begin{equation}
\tau^{-} := \frac{1}{\all{n}+m_{\min}-1} \frac{\all{v}_{\min }}{\all{v}},
\end{equation}%
where $m_{\min}$ and $\all{v}_{\min }$ are, respectively, the number of seats and the number of votes cast for all parties in the district with the fewest seats, and, if there are multiple such districts, in the one with the fewest votes.

Once we posit that the relation between renormalized vote shares and seat shares is to satisfy (\ref{eq:seatAlloc}), it is clear that the renormalized vote shares cannot be arbitrarily small. In particular, as $q_i$ is by definition non-negative, it is necessary that
\begin{equation}
    \rel{p}_i \left(1 + \frac{\rel{n}}{2m}\right) \ge \frac{1}{2m}.
\end{equation}
By transforming it, we arrive at the following condition:

\begin{condition} \label{condNT}
    If we require that $q_i > 0$, it is necessary that%
    \begin{equation}
        \rel{p}_{i}>\frac{1}{2m+\rel{n}},  \label{condRelev1}
    \end{equation}%
    where $\lfrac{1}{(2m+\rel{n})}$ is the \emph{threshold of relevance}, denoted by $t$.
\end{condition}

\noindent Note that Condition \ref{condNT} is equivalent to the condition for relevance in (\ref{eq:relevParties}), i.e., the $i$-th party is relevant if and only if it satisfies (\ref{condRelev1}). It can also be demonstrated that even if we were to modify the definition of $\rel{n}$, parties which do not satisfy (\ref{condRelev1}) would not obtain positive seat shares:

\begin{observation}
    Fix any $\tau^{*} \in [0, 1)$, and let $\Pi^{*} := \{i = 1, \dots, \all{n}: \all{p}_i > \tau^{*}\}$. Note that
    \begin{equation}
        q_i^{*} := \frac{v_i}{\sum_{j \in \Pi^{*}} v_j} \left(1 + \frac{|\Pi^{*}|}{2m}\right) - \frac{1}{2m}
    \end{equation}
    is positive for every $i \in \Pi^{*}$ and every $\mathbf{\all{p}} := (\all{p}_1, \dots, \all{p}_{n}) \in \Delta_{\all{n}}$ if and only if $\Pi^{*} \setminus \{1, \dots, \rel{n}\} = \varnothing$.
\end{observation}


\begin{remark}
Condition \ref{condNT} can also be expressed in equivalent terms of potential interest to some readers:%
\begin{equation}
\rel{n} > \frac{\sum_{j=1}^{\rel{n}} v_{j}}{v_{i}}-2m,  \label{condRelev2}
\end{equation}%
\begin{equation}
2m > \sum_{j=1}^{\rel{n}}\frac{v_{j}-v_{i}}{v_{i}}.  \label{condRelev3}
\end{equation}
\end{remark}

\begin{remark}
The threshold of relevance can be used to express (\ref{seatShare}) in yet
another form:
\begin{equation}
q_{i}=\frac{1}{2mt}(\rel{p}_{i}-t) ,  \label{seatBiasAlt}
\end{equation}%
which demonstrates that the seat shares are proportional not to the renormalized
vote shares, but only to their excess over the threshold of relevance.
\end{remark}

\begin{remark}
    Note that for the single-district case $(m=s)$, our definition of the threshold of relevance is in accord with the earlier findings about the thresholds of exclusion and inclusion. It is easy to show that (\ref{seatShare}) gives at least $1/2$ seat for a party fulfilling $\all{p}_{i} > (m+1)^{-1} \geq (2m+\all{n})^{-1}$, and at most $\frac{1}{2} (m+1)/(\all{n}+m-1) \leq 1/2$ for a party satisfying $\rel{p}_{i} < (\all{n}+m-1)^{-1}$, though the latter is not necessarily non-relevant, as it cannot be ruled out that $t (\sum_{j=1}^{\rel{n}} \rel{p}_j) < (\all{n}+m-1)^{-1}$. Given that seats are integer, a party exceeding the threshold of exclusion is guaranteed to obtain at least one, and a party below the threshold of inclusion is guaranteed to obtain none.
\end{remark}

\begin{remark}
    Note that $\tau^{-}$ is approximately of the order of $s^{-1}$. However, from (\ref{eq:seatAlloc}) we obtain:
    \begin{equation}
        q_i \rvert_{\rel{p}_i = 1/s} > 0 \textrm{\quad if and only if \quad} \frac{\rel{n}}{2m} \ge s - 1,
    \end{equation}
    which is a rather unrealistic condition, as even for $m=1$ that would require the number of relevant parties to be at least twice the number of districts minus $2$. Thus, the aggregate threshold of inclusion is usually lower than the threshold of relevance, demonstrating that \textbf{A2} is not redundant.
\end{remark}

\begin{remark}
    To the extent (\ref{eq:seatAlloc}) is an approximation of the seat allocation under the Jefferson--D'Hondt method, the threshold of relevance is an estimate of the aggregate natural threshold.
\end{remark}

\subsection{Continuity of threshold effects}

Given the discreteness of the number of relevant parties, one could prima facie expect that if $\mathbf{p}$ changes so that a previously non-relevant party's vote share exceeds the threshold of relevance or a previously relevant party's vote share no longer does, the number of seats for every other party would change discontinuously.
This would obviously constitute a significant obstacle to applying Theorem \ref{thm:main} to estimate seat allocations under circumstances where $\all{p}_{1},\dots ,\all{p}_{n}$ are known only approximately (for instance, obtained from opinion polls) and some parties are in the vicinity of the natural threshold. Fortunately, this is not the case:

\begin{proposition}
    For every $k < n$ if the first $k$ parties are relevant, then $(\all{p}_{1},\dots ,\all{p}_{k+1}) \mapsto (q_{1},\dots ,q_{k+1})$ is continuous in $\all{p}_{k+1}$ throughout $[0, \all{p}_{k}]$.
\end{proposition}

\begin{proof}
Let $t^{\prime} := (2m+k+1)^{-1}$ and $\all{p}^{\prime} := \frac{t^{\prime}}{1-t^{\prime}} \sum_{j=1}^{k} \all{p}_{j}$. Note that $\all{p}^{\prime} = t^{\prime} \left(\sum_{j=1}^{k} \all{p}_{j} + \all{p}^{\prime}\right)$ and that the $(k+1)$-th party is relevant if and only if $p_{k+1} > p^{\prime}$.

Fix $\all{p}_1, \dots, \all{p}_k$ and let $\all{p}_{k+1} = \all{p} \in [0, \all{p}_{k}]$ be variable. For $\all{p} \le \all{p}^{\prime}$ we have%
\begin{equation}
q_{i}\big\rvert_{\all{p}_{k+1} = p} =
\begin{dcases}
\frac{2m+k}{2m}\frac{\all{p}_{i}}{\sum_{j=1}^{k} \all{p}_{j}}-\frac{1}{2m} & 
\textrm{ for } i=1,\dots ,k, \\ 
0 & \textrm{ for } i = k+1, \dots, \all{n}.
\end{dcases}%
\end{equation}%
On the other hand, for $\all{p}>\all{p}^{\prime }$ and $i=1,\dots ,k,k+1$ we obtain
\begin{equation}
q_{i} \big\rvert_{\all{p}_{k+1} = p} =\frac{2m+k+1}{2m} \frac{\all{p}_{i}}{\all{p}+\sum_{j=1}^{k} \all{p}_{j}}-\frac{1}{2m}.
\end{equation}
Thus, we only need to consider the case of $\all{p} \searrow \all{p}^{\prime}$. For $i=1,\dots ,k$ we have
\begin{align}
\lim_{\all{p}\searrow\, \all{p}^{\prime}} q_{i} \big\rvert_{\all{p}_{k+1} = p} & = 
\frac{2m+k+1}{2m} \frac{\all{p}_{i}}{\all{p}^{\prime }+\sum_{j=1}^{k} \all{p}_{j}}-\frac{1}{2m} \\
& =\frac{\all{p}_{i}}{2m p^{\prime}}-\frac{1}{2m} =
\frac{1-t^{\prime }}{2mt^{\prime}} \frac{\all{p}_{i}}{\sum_{j=1}^{k}\all{p}_{j}}-\frac{1}{2m} \\
& =\frac{2m+k}{2m}\frac{\all{p}_{i}}{\sum_{j=1}^{k}\all{p}_{j}}-\frac{1}{2m},
\end{align}%
and for $i=k+1$ we obtain
\begin{equation}
\lim_{\all{p}\searrow\, \all{p}^{\prime}} q_{i} \big\rvert_{\all{p}_{k+1} = p} =
\frac{\all{p}^{\prime}}{2m \all{p}^{\prime}} -\frac{1}{2m} = 0,
\end{equation}%
as desired.
\end{proof}

\funding{Supported by the Polish National Science Center (NCN) under grant no. 2019/35/B/HS5/03949.}

\bibliographystyle{plainnat}
\bibliography{DHondt}

\end{document}